\documentclass[aps,preprint]{revtex4}
\usepackage{amsfonts}
\usepackage{amssymb,epsfig}
\usepackage{latexsym}
\usepackage{amsmath}
\usepackage{graphicx}
\begin{document}

\title{Einstein-Born-Infeld on Taub-NUT Spacetime in $2k+2$ Dimensions }

\author{A. Khodam-Mohammadi\footnote{%
Email address: \texttt{khodam@basu.ac.ir}}}

\address{Physics Department, Faculty of Science, Bu-Ali Sina University, Hamedan, Iran}

\begin{abstract}
We wish to construct solutions of Taub-NUT spacetime in
Einstein-Born-Infeld gravity in even dimensions. Since Born-Infeld
theory is a nonlinear electrodynamics theory, in leads to nonlinear
differential equations. However a proper analytical solution was not
obtain, we try to solve it numerically (by the Runge-Kotta method)
with initial conditions coinciding with those of our previous work
in Einstein-Maxwell gravity. We solve equations for 4, 6 and 8
dimensions and do data fitting by the least-squares method. For
$N=l=b=1$, the metric turns to the NUT solution only in 8
dimensions, but in 4 and 6 dimensions the spacetime does not have
any Nut solution.
\end{abstract}
\maketitle

\section{Introduction}

In the recent years, a great deal of attention has been focused on
the Born–Infeld theory. This theory emerges from string theory with
a parameter $b$ with dimension of length. The covariant action of
this field is
\begin{equation}
\frac{1}{4\pi b^{2}}\int d^{4}x\sqrt{-g}\left\{
1-\sqrt{1-\frac{1}{2}b^{2}F_{\mu \nu }F^{\mu \nu
}-\frac{1}{16}b^{4}(\varepsilon ^{\mu \nu \rho \sigma }F_{\mu \nu
}F_{\rho \sigma })^{2}}\right\}  \label{Covaction}
\end{equation}
where $F_{\mu \nu }$ is derived from a vector potential $A_{\mu }$
as $ F_{\mu \nu }=\partial _{\mu }A_{\nu }-\partial _{\nu }A_{\mu
}.$ Since this theory is a nonlinear version of electromagnetism,
the self energy of the electron in this field is finite. This fact
was a first motivation for Born and Infeld who formulated this
theory in 1934 \cite{BI}. In 1935, Hoffman \cite {Hoff} joined
general relativity to Born-Infeld electrodynamics to obtain a
spherically symmetric solution representing the gravitational field
of a charged object.

The properties of Taub-Nut spacetime are described in Refs.
\cite{HawHun, mann99, DKh}. The spherically symmetric solutions and
rotating solutions in (Einstein/Gause-Bonnet/Lovelock)-Born-Infeld
gravity with or without cosmological constant and dilaton field have
been considered by many authors \cite{demi, cai, dey, DBS, DHSR,
DSH}. In this paper, we are interested in NUT solutions of
Einstein-Born-Infeld (EBI) gravity. The thermodynamic properties of
this spacetime have been investigated in Einstein-Maxwell and
Gause-Bonnet gravity \cite{DKh, KhMo}.

The outline of my paper is as follows. A brief review of the field
equations of EBI gravity and field equations of taub-NUT spacetime
in EBI gravity are described in Section \ref{field}. Also in this
section we obtain non linear equations to be solved numerically in
Section \ref {num}. At last, the paper ends with some concluding
remarks.

\section{Field equations\label{field}}

\bigskip The gravitational action for Einstein gravity in $n+1$ dimension in
the presence of BI field with a cosmological constant $\Lambda
=-n(n-1)/2l^{2}$ is
\begin{equation}
I_{G}=-\frac{1}{16\pi }\int d^{n+1}x\sqrt{-g}\left[ R-2\Lambda +L(F)\right] +%
\frac{1}{8\pi }\int_{\partial M}d^{n}x\sqrt{-\gamma }K(\gamma ).
\label{EBIaction}
\end{equation}
The first term is the Einstein-Hilbert action and second term is the
Gibbons-Hawking boundary term which is chosen such that the
variational principle is well-defined. The manifold $\mathcal{M}$
has the metric $g_{\mu \nu }$. $K$ is trace of the extrinsic
curvature $K^{\mu \nu }$ of any boundaries $\partial \mathcal{M}$ of
the manifold $\mathcal{M}$ with the induced metric $\gamma _{ij}. $
In Eq. (\ref{EBIaction}) $L(F)$ is the Lagrangian of BI field given
as
\begin{equation}
L(F)=\frac{4}{b^{2}}\left( 1-\sqrt{1+\frac{b^{2}F^{2}}{2}}\right) .
\label{BIlagran}
\end{equation}
In these equations $F^{2}=F_{\mu \nu }F^{\mu \nu }$, where $F_{\mu
\nu }$ is the electromagnetic tensor field and $b$ is the BI
parameter. As one can see in the limit $b\rightarrow 0$, $L(F)$
Reduces to the standard Maxwell Lagrangian $-F^{2}$. As
$b\rightarrow \infty $, $L(F)\rightarrow 0$.

The Field equations are obtained By varying the action (\ref{EBIaction})
with respect to the gauge field $A_{\mu }$ and the gravitational field $%
g_{\mu \nu }$ as
\begin{equation}
\partial _{\mu }\left( \frac{\sqrt{-g}F^{\mu \nu }}{\sqrt{1+\frac{b^{2}F^{2}%
}{2}}}\right) =0,  \label{fieldA}
\end{equation}
\begin{equation}
G_{\mu \nu }+\Lambda g_{\mu \nu }=\frac{1}{2\pi }\left( g_{\mu \nu }L(F)+%
\frac{4F_{\mu \lambda }F_{\nu }^{\lambda }}{\sqrt{1+\frac{b^{2}F^{2}}{2}}}%
\right)  \label{fieldg}
\end{equation}
where $G_{\mu \nu }$ is the Einstein tensor.

The Euclidean section of the $(2k+2)$ dimensional Taub-NUT spacetime
can be written as
\begin{equation}
ds^{2}=F(r)(d\tau +N\mathcal{A})^{2}+F^{-1}(r)dr^{2}+(r^{2}-N^{2})d\Xi _%
\mathcal{B}  \label{Nutspace}
\end{equation}
where $\tau $ is the coordinate of the fibers $S^{1}$ and
$\mathcal{A}$ is the K\"{a}hler form of the base space
$\mathcal{B}$, $N$ is the NUT charge and $F(r)$ is a function of
$r.$ The metric $d\Xi _{\mathcal{B}}$ is a $2k$-dimensional base
space Einstein-K\"{a}hler manifold $\mathcal{B}.$

Here, we consider only the cases where all the factor spaces of
$\mathcal{B}$ have positive curvature. Thus, the base space
$\mathcal{B}$ may be the product of $2$-sphere $S^{2}$ and/or
$\mathbb{CP}^{k}$ spaces for all values of $k$. For completeness, we
give the $1$-forms and the metrics of these factor spaces. The
$1$-forms and the metrics of $S^{2}$ are
\begin{eqnarray}
\mathcal{A}_{i} &=&2\cos \theta _{i}d\phi _{i},  \label{Smetric} \\
d\Omega _{i}^{2} &=&d\theta _{i}^{2}+\sin ^{2}\theta _{i}d\phi
_{i}^{2}, \nonumber
\end{eqnarray}
and those of $\mathbb{CP}^{k}$ are
\begin{eqnarray}
\mathcal{A}_{k} &=&2(k+1)\sin ^{2}\xi _{k}(d\psi _{k}+\frac{1}{2k}\mathcal{A}%
_{k-1}),  \label{Ak} \\
d\Sigma _{k}^{2} &=&2(k+1)\left\{ d\xi _{k}^{2}+\sin ^{2}\xi
_{k}\cos ^{2}\xi _{k}(d\psi
_{k}+\frac{1}{2k}\mathcal{A}_{k-1})^{2}+\frac{1}{2k}\sin ^{2}\xi
_{k}d\Sigma _{k-1}^{2}\right\} \nonumber
\end{eqnarray}
where $\mathcal{A}_{k-1}$\ is the K\"{a}hler potential of
$\mathbb{CP}^{k-1}$. In Eqs. (\ref{Ak}) $\xi _{k}$\ and $\psi _{k}$\
are the extra coordinates corresponding to $\mathbb{CP}^{k}$ with
respect to $\mathbb{CP}^{k-1}$. The metric $\mathbb{CP}^{k}$ is
normalized so that the Ricci tensor is equal to the metric, $R_{\mu
\nu }=g_{\mu \nu }$. The $1$-form and the metric of
$\mathbb{CP}^{1}$ is
\begin{eqnarray}
\mathcal{A}_{1} &=&4\sin ^{2}\xi _{1}d\psi _{1}  \label{A1} \\
d{\Sigma _{1}}^{2} &=&4\left( {d\xi _{1}}^{2}+\sin ^{2}\xi _{1}\cos
^{2}\xi _{1}{d\psi _{1}}^{2}\right)  \label{CP1}
\end{eqnarray}
and those of $\mathbb{CP}^{k}$ can be constructed through the use of
Eqs. (\ref{Ak}). One can see that $\mathbb{CP}^{1}=S^{2}$ by
changing parameter $\theta =2\xi.$

The gauge potential has the form
\begin{equation}
A=g(r)(d\tau +N\mathcal{A}),  \label{A}
\end{equation}
where $g(r)$ is a function of $r$. The electromagnetic field
equation (\ref{fieldA}) for the metric (\ref{Nutspace}) with the
vector potential (\ref{A}) in $(2k+2)$ dimensions is
\begin{eqnarray}
\left[ 1+4kb^{2}h(r)^{2}N^{2}\right] (r^{2}-N^{2})h^{\prime \prime
}(r)+2krb^{2}(r^{2}-N^{2})h^{\prime }(r)^{3}  \nonumber\\
+4kb^{2}h(r)(r^{2}-N^{2})(3r^{2}-2N^{2})h^{\prime
}(r)^{2}+8k\left[ 2r^{2}+(2k+1)N^{2}\right] b^{2}h(r)^{3}  \nonumber \\
+2r\left\{ 4k[(k-1)N^{2}+3r^{2}]b^{2}h(r)^{2}+(k+2)\right\}
h^{\prime }(r)+2(2k+1)h(r)=0 \label{EqA}
\end{eqnarray}
where $h(r)=g(r)(r^{2}-N^{2})$ and prime denotes a derivative with
respect to $r$. To find the function $F(r)$ in the metric
(\ref{Nutspace}), one may use any components of Eq. (\ref{fieldg}).
The simplest equation is the $tt$ component of these equations which
can be written as
\begin{eqnarray}
krb^{2}p(r)F^{\prime }(r)+kb^{2}p(r)\frac{[(2k-1)r^{2}+N^{2}]}{%
(r^{2}-N^{2})}F(r)+2(r^{2}-N^{2})^{2}  \nonumber\\
-\left\{ \lbrack
2+k(2k+1)\frac{b^{2}}{l^{2}}](r^{2}-N^{2})+kb^{2}\right\}
p(r)+8kN^{2}b^{2}g(r)^{2}=0 \label{EqF}
\end{eqnarray}
where $p(r)$ is a function of $r$ as
\begin{equation}
p(r)=\sqrt{[g^{\prime
}(r)^{2}b^{2}+1](r^{2}-N^{2})^{2}+4kb^{2}N^{2}g(r)^{2}}. \label{EqP}
\end{equation}
Eqs. (\ref{EqA}) and (\ref{EqF}) are nonlinear. These equations must
be solved to find $F(r)$. An analytic solution of these equations
has not been found and at last we try to solve them numericallly.

\section{Numerical solutions\label{num}}

\bigskip The solutions of Eqs. (\ref{EqA} and\ \ref{EqF}) describe NUT
solutions, if

(I) $F(r=N)=0,$

(II) $g(r=N)=0,$

(III) $g^{\prime }(r=N)=S(k),$

(IV) $F(r)$ should have no positive roots at $r>N$.

The first condition comes from the fact that the metric
(\ref{Nutspace}) has a singularity at $r=N$ or in other word all
extra dimensions should collapse to zero at a fixed point set of
$\partial /\partial \tau $. The second one come from the regularity
of vector potential at $r=N$, the third one from our previous work
\cite{DKh} for Einstein-Maxwell gravity, and the fourth condition
means that $r=N$ should be the outer horizon. $S(k)$ is the value of
$g^{\prime }(N)$ in any dimension.

By using these conditions, we solve Eqs. (\ref{EqA} and\ \ref{EqF})
numerically by the following steps

1. Solve Eq. (\ref{EqA}) with the Runge-Kutta method with two
initial conditions (conditions (II) and (III) for NUT solutions).

2. By data fitting in the least squares method, an appropriate
function for $ g(r) $ is obtained. The function that we use for
fitting is $g(r)=\underset{n}{\sum }a_{n}r^{n}\times f(r)$ where
$f(r)$ is the corresponding function is obtained analytically in the
previous work on Einstein-Maxwell (EM) gravity in this spacetime
\cite{DKh}.

3. By substituting the data of $g(r)$ and $g^{\prime }(r)$ into Eq.
(\ref{EqP}), $p(r)$ is calculated. Then data fitting is done as the
previous step.

4. Insert $g(r)$ and $p(r)$ in Eq. (\ref{EqF}). By solving this
differential equation numerically, with initial condition,
conditions and data fitting, $F(r)$ is obtained.

We solve these equations for $N=l=b=1.$ The values of $S_{k}$ are

\begin{center}
$S_{k}=\left\{
\begin{array}{c}
0.5\text{ \ \ \ \ \ \ \ \ \ for }k=1\text{ (4 Dim.)} \\
-1\text{ \ \ \ \ \ \ \ \ \ for }k=2\text{ (6 Dim.)} \\
1.25\text{ \ \ \ \ \ \ \ \ for }k=3\text{ (8 Dim.)}
\end{array}
\right. $
\end{center}

In figures 1-5, the function $F(r)$ is plotted for 4, 6 and 8
dimensions. as we see in these figures, in 4D, there are three roots
in the region $r\geq N$ at $r=1$, $ 1.12$ and $2.265$. In 6D, there
are two roots in the region $r\geq N$ at $r=1$, and $2.318$. Since
the fourth condition is not satisfied, therefore, the metric is no
NUT solution in these dimensions. But in 8D, as we see in Fig.
\ref{f8d}, there are no roots greater than $N$. Therefore the metric
has a Nut solution.
\begin{figure}[tbp]
\epsfxsize=6cm \centerline{\epsffile{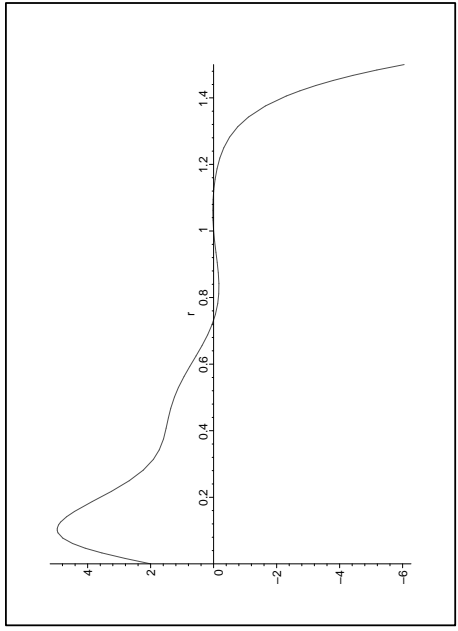}}\caption{F(r) in 4
dimensions ($0<r<1.5$)} \label{f4d1}
\end{figure}
\begin{figure}[tbp]
\epsfxsize=6cm \centerline{\epsffile{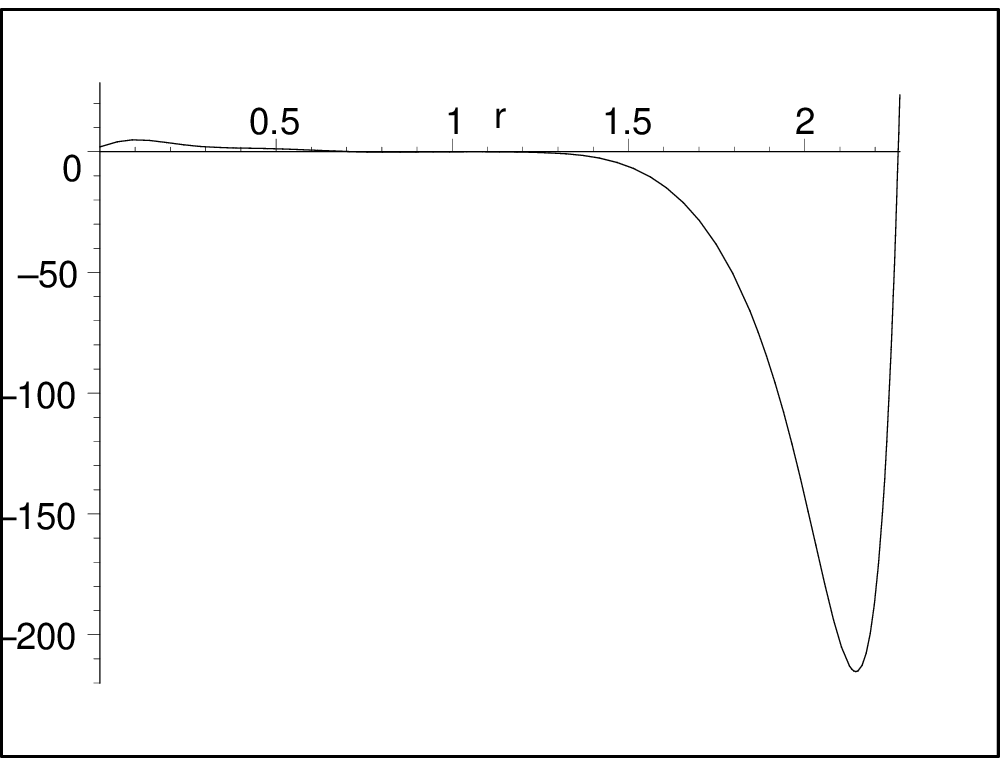}}\caption{F(r) in 4
dimensions ($0<r<2.3$)} \label{f4d2}
\end{figure}
\begin{figure}[tbp]
\epsfxsize=6cm \centerline{\epsffile{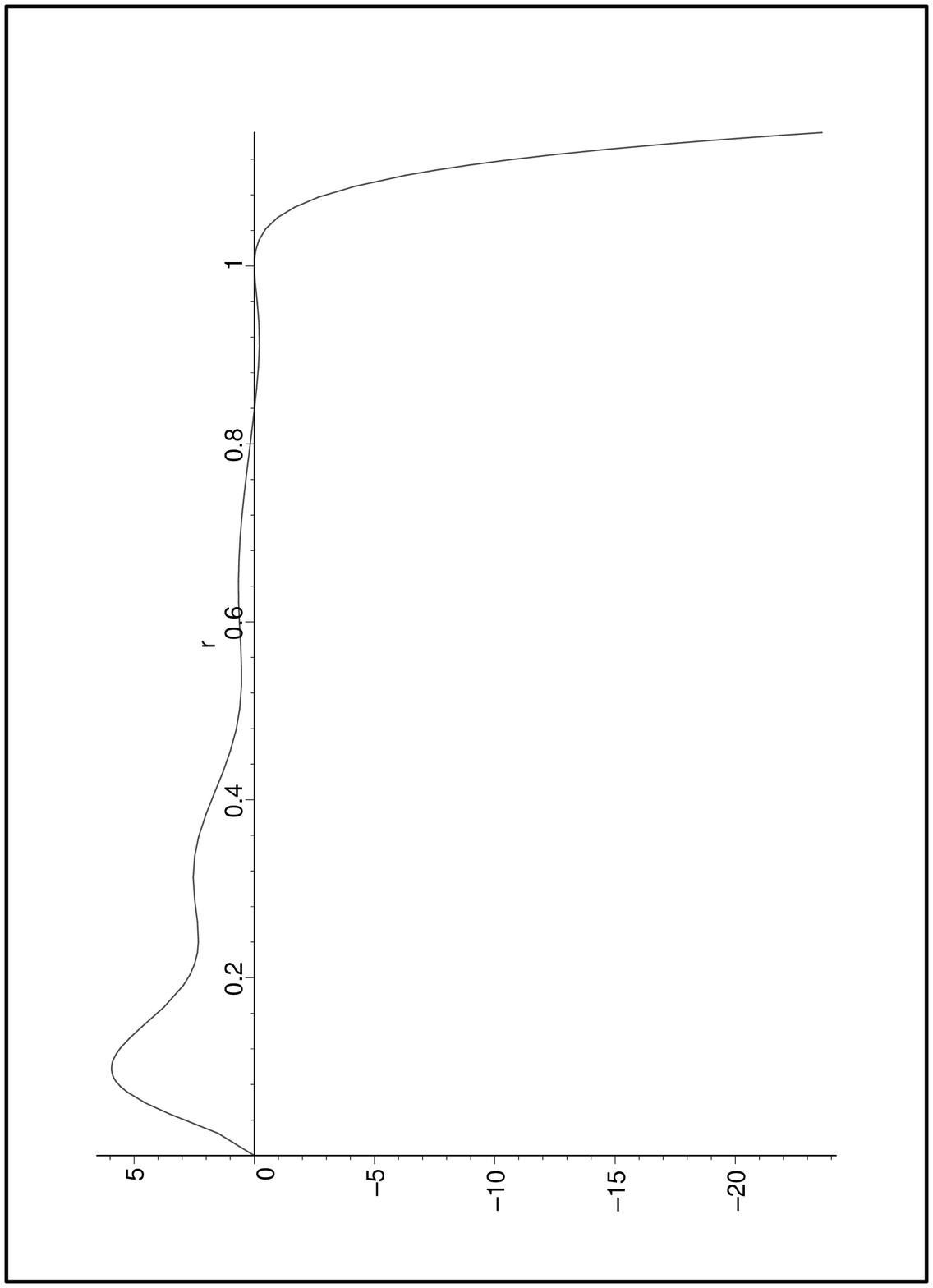}}\caption{F(r) in 6
dimensions ($0<r<1.15$)} \label{f6d1}
\end{figure}
\begin{figure}[tbp]
\epsfxsize=6cm \centerline{\epsffile{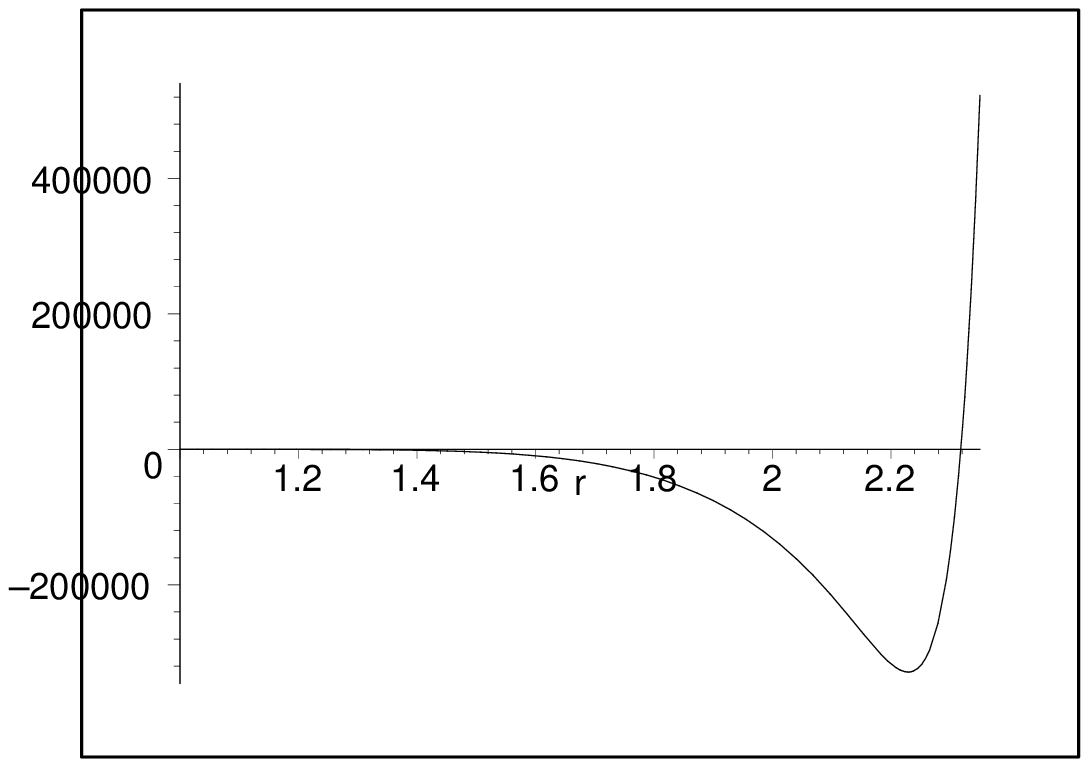}}\caption{F(r) in 6
dimensions ($0<r<2.3$)} \label{f6d2}
\end{figure}
\begin{figure}[tbp]
\epsfxsize=6cm \centerline{\epsffile{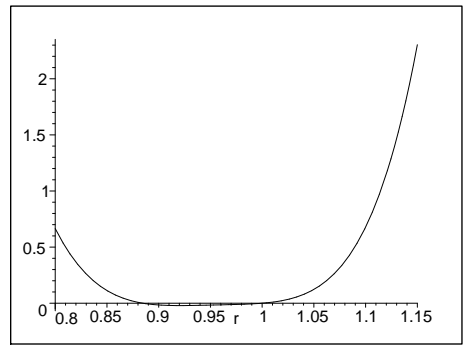}}\caption{F(r)in 8
dimensions ($0<r<1.15$)} \label{f8d}
\end{figure}

\section{Conclusion}

The Born-Infeld theory on Taub-NUT spacetimes were investigated. Two
field equations were driven after varying the gravitational action.
We fined two nonlinear differential equations for Taub-NUT spacetime
in $2k+2$ dimensions. A Numerical solution for these equations was
suggested. By having conditions for NUT solutions and for $N=l=b=1$,
we saw that there is no NUT solution for 4 and 6 dimensions. In 8
dimensions, there exist NUT solutions. As one can see, These
solutions are checked only for $ N=b=1 $. In the future, if these
equations are solved for all values of `$N$' and `$b$', we may
obtain regions of NUT solutions in any dimensions.


\begin{thebibliography}{99}

\bibitem{BI} M. Born and L. Infeld, Proc. Roy. Soc. Lond. \textbf{A} \textbf{
144}, 425 (1934).

\bibitem{Hoff} B. Hoffmann, Phys. Rev. \textbf{47}, 877 (1935).

\bibitem{HawHun} S. W. Hawking, C. J. Hunter and D. N. Page, Phys. Rev. D
\textbf{59}, 044033 (1999).

\bibitem{mann99} R. B. Mann, Phys. Rev. D \textbf{60}, 104047 (1999).

\bibitem{DKh} M. H. Dehghani and A. Khodam-Mohammadi; phys. Rev. D. \textbf{
73}, 124039 (2006).

\bibitem{demi} M. Demianski, Found. Phys. \textbf{16}, (1986) 187; D.
Wiltshire, Phys. Rev. D \textbf{38}, 2445 (1988).

\bibitem{cai} R. G. Cai, D. W. Pang and A. Wang, Phys. Rev. D \textbf{70}, 124034
(2004).

\bibitem{dey} T. K. Dey, Phys. Lett. B \textbf{595 }, 484 (2004).

\bibitem{DBS} M. H. Dehghani, G. H. Bordbar and M. Shamirzaei, Phys. Rev. D
\textbf{74}, 064023 (2006).

\bibitem{DHSR} M. H. Dehghani, S. H. Hendi, A. Sheykhi, H. Rastegar Sedehi,
JCAP \textbf{0702}, 020 (2007).

\bibitem{DSH} M. H. Dehghani, A. Sheykhi, S. H. Hendi, Phys. Lett. \textbf{B}
\textbf{659}, 476 (2008).

\bibitem{KhMo} A. Khodam-Mohammadi and M. Monshizadeh; phys. Rev. D. \textbf{
79}, 044002 (2009).

\end{thebibliography}
\end{document}